\documentclass[epj]{svjour}

\usepackage{epsfig}
\usepackage{latexsym}

\begin{document}

\title{Analytic solution of a static scale-free network model}

\author{Michele Catanzaro \and Romualdo Pastor-Satorras}

\institute{Departament de F\'\i sica i Enginyeria Nuclear, Universitat
  Polit\`ecnica de Catalunya, Campus Nord B4, 08034 Barcelona, Spain}

\date{\today}

\abstract{% 
  We present a detailed analytical study of a paradigmatic scale-free
  network model, the \emph{Static Model}. Analytical expressions for
  its main properties are derived by using the \emph{hidden variables}
  formalism. We map the model into a canonic hidden variables one, and
  solve the latter. The good agreement between our predictions and
  extensive simulations of the original model suggests that the
  mapping is exact in the infinite network size limit. One of the most
  remarkable findings of this study is the presence of relevant
  disassortative correlations, which are induced by the physical
  condition of absence of self and multiple connections.
}

\maketitle

\section{Introduction}
\label{sec:introduction}

In the last few years a considerable amount of research effort has
been devoted to the study of a large array of natural and man-made
systems that can be described in terms of networks. In fact, systems
as diverse as the Internet \cite{romuvespibook}, the World-Wide Web
\cite{hubbook}, collaborations networks \cite{newman01a,schubert}, the
web of sexual contacts \cite{amaral01}, foodwebs
\cite{montoya02,DiegoGuido}, protein interactions networks
\cite{wagner01}, metabolic networks \cite{Jeong00}, and many others
can be represented, at a certain level of approximation, as networks
or graphs \cite{bollobas98}, in which vertices represent the
elementary units composing the system, while the edges stand for the
relations or interactions present between pairs of elements.  These
kind of systems were in the previous century the subject of study of
classical graph theory. Recently, however, the availability of large
data sets and more powerful computer resources, together with the
application of new statistical tools, has led to the development of a
modern theory of complex networks \cite{barabasi02,mendesbook}, which
is nowadays one of the most active fields in the statistical physics
of complex systems.

Despite of the wide variety of the systems considered in this field,
some common characteristics seems to be present in almost all complex
networks. Among them, the most remarkable is probably the fact that
many real-world networks exhibit a fat-tailed degree distribution
$P(k)$.  That is, the probability that a randomly chosen vertex has a
number of emerging edges equal to an integer $k$ (the degree of the
vertex) has the form for large $k$
\begin{equation}
  P(k) \sim k^{-\gamma},
\end{equation} 
where the degree exponent $\gamma$ commonly ranges in the interval $\gamma \in
[2,3]$ \cite{mendesbook}.  This suggest the presence of a
heterogeneous hierarchy of vertices, lacking a characteristic degree
value, which has result in the common denomination of scale-free
networks \cite{barab99}.  Moreover, the presence of a scale-free
degree distribution implies that the degree fluctuations are unbounded
in the infinite network size limit, i.e. $\langle k^2 \rangle \to\infty$ when $N \to\infty$,
which has a considerable impact on the behavior of dynamical processes
taking place on top of the network. For instance, it has been shown
that scale-free networks are extremely resilient to random damage
\cite{jeong00,newman00,havlin01}, while at the same time they are very
weak in front to the spread of epidemic processes
\cite{pv01a,lloyd01}.

In addition to the degree distribution, it has been po\-in\-t\-ed out that
real networks are further characterized by the presence of degree
correlations, which translate in the fact that the degrees at the end
points of any given edge are not usually independent. This kind of
degree correlations can be quantitatively expressed in terms of the
conditional probability $P(k' | k)$ that a vertex of degree $k$ is
connected to a vertex of degree $k'$ \cite{knnromu,hiddenromu}. From a
numerical point of view, the presence of correlations can be
conveniently studied by means of the average degree of the nearest
neighbors of the vertices of degree $k$, formally defined as
\cite{knnromu}
\begin{equation}
  \bar{k}_{nn}(k) = \sum_{k'} k' P(k' | k).
\end{equation}
A first classification of networks has been proposed according to the
nature of their correlations \cite{assortative}.  Thus, when
$\bar{k}_{nn}(k)$ is a growing function of $k$, the network is said to
exhibit \textit{assortative mixing}, while a decreasing
$\bar{k}_{nn}(k)$ function is typical of \textit{disassortative
  mixing}.

The appealing evidence for the existence of scale-free networks has
prompted the development of numerous models, aimed at understanding
the origin of fat-tailed degree distributions, or even the nature of
degree correlations \cite{mio}. Scale-free network models can be
roughly divided in two main classes: \textit{Growing network models},
capitalizing in the original Barab\'asi-Albert model \cite{barab99},
focus their approach on the evolution of the network, rather than on
its structure. The key ingredient of these models consist in
considering the network as a result of a growth process, in which new
vertices and edges are sequentially added to the system following a
prescribed set of dynamical rules (usually inspired in the
preferential attachment or \textit{rich-get-richer} paradigm
\cite{barab99}). On the other hand, \textit{static network models}
consider networks with a constant number of vertices $N$, among which
edges are drawn following different probabilistic rules. In this
sense, Ref.~\cite{fitnesguido} proposed for the first time a static
network model yielding a scale-free degree distribution, while
Ref.~\cite{hiddenromu} (see also Ref.~\cite{fitnessoder}) proposed a
general class of static network models, the \textit{hidden variables
  network models}, which allow to develop a systematic analytical
formalism for this class of systems.

One of the models belonging to the class of static network models that
has recently attracted some attention, is the so-called \textit{static
  model} (SM), recently proposed in Ref.~\cite{static}. The interest
raised by the SM has a twofold origin. Firstly, its definition is very
simple, and allows to generate large networks with any desired
degree exponent with a reasonable amount of computer effort. Secondly,
its use as a benchmark to check the properties of both scale-free
networks and dynamical processes running on top of networks has become
quite widespread
lately~\cite{static,barthel1,barthel2,sandpile1,sync}.  In spite of
this wide interest and use, however, little is known from an
analytical point of view about the properties of the networks
generated by the SM (apart from its scale-free nature), especially in
what refers the nature of its possible degree correlations.

In this paper we present an analytical treatment of the SM based in a
mapping to a hidden variables network model. Using the hidden
variables formalism, we are able to provide analytic expressions for
the main properties of the SM, in particular for the degree
distribution and the degree correlations, as measured by the
$\bar{k}_{nn}(k)$ function, 
showing a very good agreement with extensive numerical simulations of
the original SM model. One of the most remarkable findings that we
report is the presence of strong disassortative degree correlations in
the SM for values of the degree exponents close to $2$, in agreement
with the theoretical arguments put forward in
Ref.~\cite{mariancutofss}. The presence of this correlations indicate
that the results of dynamical processes running on top of networks
generated with the SM should be interpreted with great care, in order
to discern the effects due to the scale-free nature of the networks
from those related with the presence of intrinsic degree correlations.

The paper is organized as follows.  In Section~\ref{sec:model} we
review the definition of the SM model, as well as some of the its
properties, that can be derived by using simple qualitative arguments.
In Sec.~\ref{sec:mapp-hidd-vari} we provide an overview of the general
formalism for hidden variables network models and discuss how can we
map the original SM into a model belonging this class of networks. In
Sec.~\ref{sec:analytic-solution} we proceed to solve the mapped model,
and provide analytical expression for its main quantitative
properties. The analytical results obtained are checked by means of
direct numerical simulations of the original SM in
Sec.~\ref{sec:numer-simul}.  Finally, in Sec.\ref{sec:conclusions} we
draw the conclusions of our work.

\section{The static model}
\label{sec:model}

The static model (SM) was introduced in Ref.~\cite{static} as an
algorithm to generate scale-free static (i.e. not growing) networks
with any desired degree exponent $\gamma$ larger than or equal to $2$.  The
model is defined as follows: We start from $N$ disconnected vertices,
each one of them indexed by an integer number $i$, taking the values
$i=1,\ldots N$. To each vertex, a normalized probability $p_i$ is assigned,
given as function of the index $i$ by
\begin{equation}
  p_i = \frac{i^{-\alpha}}{\sum_{j=1}^N j^{-\alpha}},
  \label{eq:4}
\end{equation}
where $\alpha$ is a real number in the range $\alpha \in [0,1]$. The network is
constructed iterating the following rules: Two different vertices $i$
and $j$ are randomly selected from the set of $N$ vertices, with
probability $p_i$ and $p_j$, respectively. If there exists an edge
between these two vertices, they are discarded and a new pair is
randomly drawn.  Otherwise, an edge is created between vertices $i$
and $j$. This process is repeated until $E=m N$ edges are created in
the network, accounting for a fixed average degree $\langle k \rangle = 2E /N =
2m$.

This algorithm generates networks in which, by construction, there are
no self-connections (a vertex joined to itself) not multiple
connections (two vertices connected by more than one edge). The
corresponding degree distribution can be estimated by means of a
simple mean-field argument \cite{static}. Since edges are connected to
vertices with a probability given by the factor $p_i$, we have that
the probability that any edge belongs to the vertex $i$, with degree
$k_i$, is given by
\begin{equation}
  \frac{k_i}{\sum_j k_j} \sim p_i.
  \label{eq:1}
\end{equation}
In the large $N$ limit, approximating sums by integrals, we have that,
for $0< \alpha<1$,
\begin{equation}
  \sum_{j=1}^N j^{-\alpha} \sim \int_1^\infty j^{-\alpha} dj \sim \frac{N^{1-\alpha}}{1-\alpha}.
\end{equation}
Therefore, since $\sum_j k_j = \langle k \rangle N$, we have from Eq.~(\ref{eq:1}) that
\begin{equation}
  k_i \sim p_i \sum_j k_j  \sim 2 m (1-\alpha) \left(\frac{i}{N}\right)^{-\alpha}.
  \label{eq:2}
\end{equation}
From this last expression, and using general arguments from network
theory \cite{mendesbook}, we conclude that the degree distribution
characterizing these networks has a scale-free form, $P(k) \sim k^{-\gamma}$,
with a degree exponent 
\begin{equation}
  \gamma = 1 + \frac{1}{\alpha}.
\label{eq:degexp}
\end{equation}
Thus, tuning the parameter $\alpha$ in the range $[0,1]$ it is possible to
generate networks with a degree exponent in the range $\gamma \in [2, \infty]$. 

Just at this stage, it is possible to notice that the SM
generates networks with built-in degree correlations. From
Eq.~(\ref{eq:2}), we observe that the maximum degree, corresponding to
the index $i=1$, is given by
\begin{equation}
  k_{i=1} \sim 2m(1-\alpha) N^\alpha.
\end{equation}
This implies that the cut-off (or maximum expected degree) $k_c(N)$ in
the network \cite{dorogorev} scales with the network size as $k_c(N) \sim
N^\alpha$. Now, it has been proved that, in order to have no correlations
in the absence of multiple and self-connections, a scale-free networks
with size $N$ must have a cut-off scaling at most as $k_s(N) \sim
N^{1/2}$ (the so-called structural cut-off) \cite{mariancutofss}.
Therefore, the SM should yield correlated networks for
values $\alpha > 1/2$, i.e., for degree exponents in the interval $2<\gamma<3$,
which correspond to those values empirically observed in real
scale-free networks. In the following Sections we will provide an
analytical description of the origin and form of these degree
correlations.

\section{Mapping to a hidden variables network model}
\label{sec:mapp-hidd-vari}

In order to solve analytically the SM, it is useful to map it to a
hidden variables network model
\cite{hiddenromu,fitnesguido,fitnessoder}.  Hidden variables network
models are a generalization of the Erd\"os-R\'enyi model \cite{erdos59} in
which vertices are assigned a tag (or hidden variable) whose
statistical properties completely determine the topological structure
of the ensuing networks.

\subsection{General network models with hidden variables}
\label{sec:general-models-with}

The class of network models with hidden variables is defined as
follows \cite{hiddenromu}: Starting from a set of $N$ disconnected
vertices and a general hidden variable $h$, that can be a natural or
real number, we construct an undirected network with no self nor
multiple connections, by applying these two rules:
\begin{enumerate}
\item To each vertex $i$, a variable $h_i$ is assigned, drawn at
  random from the probability distribution $\rho(h)$.
\item For each pair of vertices $i$ and $j$, with hidden variables
  $h_i$ and $h_j$, respectively, an edge is created with probability
  $r(h_i, h_j)$ (the connection probability), where $r(h, h') \geq 0$ is
  a symmetric function of $h$ and $h'$.
\end{enumerate}

In this class of models, the degree distribution is given by 
\begin{equation}
  P(k)=\sum_h g(k|h)\rho(h)
\label{eq:pk}
\end{equation}
where the \emph{propagator} $g(k|h)$ gives the conditional probability
that a vertex with hidden variable $h$ ends up connected to $k$
vertices. The propagator is a normalized function, $\sum_k g(k|h) =1$,
whose generating function $\hat{g}(z|h)$, defined by
\begin{equation}
  \hat{g}(z|h) = \sum_k z^k g(k|h),
\end{equation}
fulfills in the general case the expression \cite{hiddenromu}
\begin{equation}
  \ln   \hat{h}(z|h) = N \sum_{h'} \rho(h') \ln \left[ 1-(1-z) r(h, h')
  \right].
  \label{eq:3} 
\end{equation}
Given the probabilities $\rho(h)$ and $r(h,h')$, Eq.~(\ref{eq:3}) must be
solved and inverted in order to obtain the corresponding propagator
and the degree distribution. Without solving this equation, however,
we can still obtain some information on the connectivity properties of
the network. Noticing that the first moment of $g(k|h)$ is given by
the first derivative of $\hat{g}(z|h)$, evaluated at $z=1$, we that
the average degree of the vertices of hidden variable $h$,
$\bar{k}(h)$, is given by
\begin{equation}
  \bar{k}(h) = \sum_k k g(k|h) = N\sum_{h'} \rho(h') r(h, h'),
  \label{eq:5}
\end{equation}
while the average degree takes the form
\begin{equation}
  \langle k\rangle = \sum_k P(k) = \sum_h \rho(h)  \bar{k}(h).
  \label{eq:6}
\end{equation}

In order to characterize degree correlations in a general model with
hidden variables, we need to provide an expression for the average
degree of the neighbors of the vertices of degree $k$,
$\bar{k}_{nn}(k)$. Consider first the average degree of the neighbors
of the vertices of hidden variable $h$, $\bar{k}_{nn}(h)$. This
quantity can be expressed as
\begin{equation}
  \bar{k}_{nn}(h) = \sum_{h'}  \bar{k}(h') p(h'|h),
\end{equation}
where $p(h'|h)$ is the conditional probability that a vertex of hidden
variable $h$ is connected to a vertex of hidden variable $h'$. To
compute this last quantity, we observe that the probability of drawing
an edge from $h$ to $h'$ is proportional to the probability of finding
an $h'$ vertex, times the probability of creating an actual edge.
Therefore,
\begin{equation}
  p(h'|h) = \frac{\rho(h')r(h, h')}{\sum_{h''} \rho(h'') r(h, h'')} = \frac{N
    \rho(h') r(h, h')}{\bar{k}(h)}.
\end{equation}
Thus, we have that
\begin{equation}
  \bar{k}_{nn}(h) = \frac{N}{\bar{k}(h)}  \sum_{h'}  \rho(h') \bar{k}(h')
  r(h,h').
\end{equation}
Finally, the correlation function $\bar{k}_{nn}(k)$ can be shown to be
given by \cite{hiddenromu}
\begin{equation}
  \bar{k}_{nn}(k) = 1+ \frac{1}{P(k)} \sum_h \rho(h) g(k|h)
  \bar{k}_{nn}(h).
  \label{eq:13}
\end{equation}

\subsection{Mapping the static model}
\label{sec:mapping-static-model}

In order to map the SM into a hidden variables network model we need
to provide a proper definition of the hidden variables $h$, their
probability distribution $\rho(h)$, and the connection probability $r(h,
h')$. A natural choice for the hidden variable is the index $i$
associated to each vertex. On its turn, the connection probability
$r(i, j)$ can be defined as the probability that vertices $i$ and $j$
end up connected in the final network.  With the original definition
of the SM, it is difficult to estimate this connection probability. In
order to overcome this difficulty, we will consider a small variation
of the algorithm defining the model. Within the original definition,
in a first step of the model, a \textit{potential} edge $(i, j)$ is
selected, by randomly choosing a pair of vertices $i$ and
$j$, with probabilities $p_i$ and $p_j$, respectively, as given by
Eq.~(\ref{eq:4}). In a second step, the potential edge $(i, j)$ is
actually created if it did not exist previously, and this process is
repeated until a given number of actual edges $E = m N$ is reached,
leading to a constant average degree $\langle k \rangle = 2m$. Thus, we can
consider this as a \textit{microcanonical} model, since the average
degree is held fixed.  This fact is in opposition with the spirit of
hidden variables network models, in which the average degree is not
constant, but tends to an asymptotic value for large network sizes
\cite{hiddenromu}. We can place the SM within this network class by
converting it to a \textit{canonical} model, in which a fixed number
$E =m N$ of potential edges is chosen, and afterwards checked for
their actual addition to the network. This canonical version of the SM
will lead to a network with a number of edges smaller than or equal to
$E$, and therefore to an average degree $\langle k \rangle \leq 2m$. However, we
expect that this canonical version of the SM will coincide with the
microcanonical original SM in the infinite network size limit, and to
observe $\langle k \rangle \to 2 m$ in the limit $N \to \infty$. The good agreement between
theoretical predictions derived from the first and simulations of the
second will confirm this claim.

Let us look at the edge creation process in the canonical version of
the SM. If we allow for the possibility to choose a potential edge
with $i=j$ (self-connection), the probability of selecting the
potential edge $(i, j)$ is $2 p_i p_j$ if $i\neq j$, and $p_i^2$ if
$i=j$. In a more compact form, the probability of choosing the
potential edge $(i,j)$ is
\begin{equation}
  \pi(i,j) = (2-\delta_{i j}) p_i p_j,
\end{equation}
where $\delta_{i j}$ is the Kronecker symbol. This probability is naturally
normalized: If we sum $ \pi(i,j)$ over all the $N(N+1)/2$ possible
potential edges (including self-connections), we have
\begin{eqnarray}
  \lefteqn{\sum_{i\leq j} \pi(i, j) =  \sum_{i\leq j} 2 p_i p_j - \sum_i p_i^2 }
  \nonumber \\
  && = \sum_{i< j} 2 p_i
  p_j + \sum_i p_i^2 = \left(\sum_i p_i \right)^2 =1,
\end{eqnarray}
since the original distribution $p_i$ is normalized. 

The probability that, in the final network, the vertices $i$ and $j$
are connected is equal to the probability that the potential edge $(i,
j)$ has been selected at least once, which is the complementary
probability that it has not been selected in the $E$ trials made to
generate the network. Therefore, for the canonical version of the
SM, we have that the probability that vertices $i$ and $j$
are connected in the network is
\begin{equation}
  p_c(i, j) = 1 - \left[1-\pi(i,j)\right]^E.
\end{equation}
This expression can be further simplified by taking the limit of large
$N$. We have that $E= mN$ and $p_i \simeq i^{-\alpha} (1-\alpha) N^{\alpha-1}$. Therefore,
we can write
\begin{equation}
  p_c(i, j) = 1 - \left[1- (2-\delta_{i j})(1-\alpha)^2 N^{2\alpha-2} i^{-\alpha} j^{-\alpha} \right]^{m
    N}.
\end{equation}
Thus, in the limit  $N\to\infty$, we can approximate this expression by an
exponential, that yields the final result
\begin{equation}
   p_c(i, j) = 1 - \exp \left[- (2-\delta_{i j})m (1-\alpha)^2 N^{2\alpha-1} i^{-\alpha} j^{-\alpha} \right].
\end{equation}
This is the probability that two vertices end up connected in the
final network in the canonical version of the SM. Therefore, in the
hidden variables version of the model we can set the connection
probability 
\begin{equation}
  r(i, j) = 1 - \exp \left[- 2 m (1-\alpha)^2 N^{2\alpha-1} i^{-\alpha} j^{-\alpha} \right],
\end{equation}
where we have neglected the Kronecker symbol, since in hidden variable
models we do no allow for the possibility of self-connections.  A
first conclusion can be extracted from this connections probability:
it does not factorize in two independent functions of $i$ and $j$.
Therefore, degree correlations will be present in the model
\cite{hiddenromu}.

To complete the mapping, we finally need to give a prescription for
the probability $\rho(i)$ of a vertex having hidden variable (index) $i$.
In the original definition of the model, the index is assigned
deterministically to each vertex. Here we will assume an approximation
already made for other models \cite{hiddenromu}, that consists in
considering the hidden variable $i$ randomly assigned from the set $\{
1, 2 , \ldots , N\}$, with probability $\rho(i)=1/N$. As we will see in the
next Sections, this assumption does not have a strong influence in
most of the analytic results, when compared with numerical simulations
of the original SM.

\section{Analytic solution}
\label{sec:analytic-solution}

\subsection{Average degree}

Let us consider in the first place the behavior of the overall average
degree, and the average degree of the vertices with index $i$. From
Eq.~(\ref{eq:5}), togheter with the definition of the probabilities
$\rho(i)$ and $r(i, j)$, we have that
\begin{eqnarray}
 \lefteqn{ \bar{k}(i) = N \sum_{j=1}^N \rho(j) r(i,j) }\nonumber \\
 &&= \sum_{j=1}^N \left\{ 1-
    \exp\left[-2m (1-\alpha)^2 N^{2\alpha-1} i^{-\alpha} j^{-\alpha} \right] \right\}.
\end{eqnarray}
Approximating sums by integrals, and performing the change of
variables $j = N x$, we are led to the expression
\begin{equation}
   \bar{k}(i) = N \int_{N^{-1}}^1 dx  \left\{ 1-
    \exp\left[-2m (1-\alpha)^2 N^{\alpha-1} i^{-\alpha} x^{-\alpha} \right] \right\}.
\end{equation}
Since $\alpha<1$, the argument of the exponential is a decreasing function
of $N$. Therefore, in the limit $N\to\infty$, we can perform a Taylor
expansion of the integrand, and approximate
\begin{eqnarray}
   \bar{k}(i) &=& N \int_{N^{-1}}^1 dx  \left[ 2m (1-\alpha)^2 N^{\alpha-1} i^{-\alpha}
     x^{-\alpha} \right]\\
   &=& 2 m (1-\alpha) \left(\frac{i}{N}\right)^{-\alpha} (1-N^{\alpha-1}).\label{eq:9}
\end{eqnarray}
For large $N$, the last term in this expression tends to $1$, and we
recover the mean-field result obtained previously for the SM,
Eq.~(\ref{eq:2}).

As for the average degree, we have from Eq.~(\ref{eq:6})
\begin{eqnarray}
  \langle k \rangle &=& \sum_{i=1}^N \rho(i) \bar{k}(i) \nonumber \\
  &=& \frac{1}{N} \sum_{i=1}^N 2 m
  (1-\alpha)\left(\frac{i}{N}\right)^{-\alpha} (1-N^{\alpha-1}) \nonumber \\
  &=& 2 m (1-N^{\alpha-1})^2, 
\end{eqnarray}
where again we have approximated sums by integrals. We observe that,
for any finite network size, $\langle k \rangle < 2m$. However, in the
thermodynamic limit $N\to\infty$, we recover the fixed degree exponent $\langle k \rangle
= 2m$ imposed by the SM.

\subsection{Degree distribution}

In order to compute the degree distribution, we must first solve
Eq.~(\ref{eq:3}) for the generating function of the propagator,
$\hat{g}(z|i)$. For the probabilities $\rho(i)$ and $r(i,j)$ we are
considering, approximating sums by integrals and performing again the
change of variables $j = Nx$, we have that
\begin{eqnarray}
  \lefteqn{\ln   \hat{g}(z|i) = N \int_{N^{-1}}^1  dx  \ln \left[
      1-(1-z) \times \right.} \\ 
    &&\times \left.\left( 1- \exp
      \left\{ -2 m (1-\alpha)^2 N^{\alpha-1} i^{-\alpha} x^{-\alpha}   \right\}  \right)  \right].
\end{eqnarray}
For $\alpha<1$, the argument in the exponential is again decreasing in the
large $N$ limit. Therefore, expanding to first order the exponential,
and then the logarithm inside the integral,
we are led to
\begin{eqnarray}
  \ln   \hat{g}(z|i) &=& N \int_{N^{-1}}^1  dx (1-z) 2 m (1-\alpha)^2 N^{\alpha-1}
  i^{-\alpha} x^{-\alpha} \nonumber \\
  &=& (1-z) \bar{k}(i). \label{eq:7}
\end{eqnarray}
Given Eq.~(\ref{eq:7}), we find that the propagator is finally given
by a Poisson form:
\begin{equation}
g(k|i)=\frac{\exp[-\bar{k}(i)]\bar{k}(i)^k}{k!}.
\end{equation}

Knowing the form of the propagator, we can derive the degree
distribution applying Eq.~(\ref{eq:pk}), i.e.
\begin{eqnarray}
  P(k) &=& \sum_{i=1}^N \rho(i) g(k|i)   = \frac{[2 m (1-\alpha)]^k} {\alpha \Gamma(k+1)} \times
  \nonumber   \\ 
  &\times& \int_1^{N^\alpha} d x \exp[-2 m (1-\alpha) x] x^{k -1 -1/ \alpha}, \label{eq:8}
\end{eqnarray}
where we have approximated $i$ as a continuous variable, performed the
change of variables $i=N x^{-1/ \alpha}$, and expressed $k! = \Gamma(k+1)$,
where $\Gamma(z)$ is the standard Gamma function. The only dependence of
expression on the network size is through the upper limit in the
integral.  Therefore, in the thermodynamic limit we can use the result
\cite{abramovitz}
\begin{equation}
  \int_1^\infty e^{-By}y^Ady=B^{-1-A}\Gamma (1+A,B),
\label{eq:defgamma}
\end{equation} 
where $\Gamma(z,a)$ is the incomplete Gamma function, to obtain
\begin{equation}
  P(k) = \frac{[2 m (1-\alpha)]^{1/ \alpha}}{\alpha} \frac{\Gamma(k-1/ \alpha,
    2m[1-\alpha])}{\Gamma(k+1)}. \label{eq:10}
\end{equation}

In order to obtain the asymptotic behavior of the degree distribution
for large $k$, we note that $\Gamma(z,a) \to \Gamma(z)$ for $z\to\infty$. Therefore, for
large $k$
\begin{equation}
  P(k) \simeq  \frac{[2 m (1-\alpha)]^{1/ \alpha}}{\alpha} \frac{\Gamma(k-1/ \alpha)}{\Gamma(k+1)} \sim
  k^{-1 -1/ \alpha}. 
\end{equation}

That is, we recover a scale-free degree distribution with a degree
exponent $\gamma = 1 + 1/ \alpha$, as derived by mean-field arguments for the
original SM.

\subsection{Degree correlations}

Next, we aim to calculate the average nearest neighbor degree of the
vertices with degree $k$, $\bar{k}_{nn}(k)$, in order to evaluate
correlations. To do so, we first compute the average nearest neighbor
degree of the vertices with index $i$, $\bar{k}_{nn}(i)$, that is
given by Eq.~(\ref{eq:5}). Using the expression for $\bar{k}(i)$ that
we have evaluated in Eq.~(\ref{eq:9}) in the large $N$ limit, we have
\begin{equation}
  \bar{k}_{nn}(i) = i^{\alpha} \sum_{j=1}^N j^{-\alpha} \left[1-\exp\left\{ -2 m
      (1-\alpha)^2 N^{2\alpha-1} i^{-\alpha} j^{-\alpha} \right\} \right].\label{eq:12}
\end{equation}
We can proceed as usual, replacing sums by integrals. In this case,
however, it is not possible to Taylor expand the integral after an
appropriate change of variables, since the extra factor $j^{-\alpha}$ in
the integral causes it to diverge in its lower limit. We must
therefore keep the full exponential form. After some formal
manipulations, we can write
\begin{eqnarray}
  \lefteqn{\bar{k}_{nn}(i) = \frac{i^\alpha (N^{1-\alpha}-1)}{1-\alpha}}  \nonumber\\
  &&+ \frac{i^\alpha}{\alpha} \int_1^\infty dx x^{-1/ \alpha} \exp\left\{ -2 m (1-\alpha)^2
    N^{2\alpha-1} i^{-\alpha} x \right\} \nonumber\\
  &&- \frac{i^\alpha N^{1-\alpha}}{\alpha} \int_1^\infty dx x^{-1/ \alpha} \exp\left\{ -2 m (1-\alpha)^2
    N^{\alpha-1} i^{-\alpha} x \right\}. \nonumber
\end{eqnarray}
After applying the identity Eq.~(\ref{eq:defgamma}), we are led to the
solution
\begin{eqnarray}
  \bar{k}_{nn}(i) &=& \frac{i^\alpha (N^{1-\alpha}-1)}{1-\alpha}  \nonumber\\
  &+& \frac{i^\alpha N^{1-\alpha}}{\alpha}[2 m (1-\alpha)^2  N^{2\alpha-1} i^{-\alpha}]^{-1+1/ \alpha} \times
  \nonumber \\
  &\times& \left\{ \Gamma(1-\frac{1}{\alpha}, 2 m (1-\alpha)^2  N^{2\alpha-1}
    i^{-\alpha})\right. \nonumber \\
  &-&  \left.\Gamma(1-\frac{1}{\alpha}, 2 m (1-\alpha)^2  N^{\alpha-1} i^{-\alpha}) \right\}.\label{eq:11}
\end{eqnarray}

As we will see in the following Section, the approximation given by
Eq.~(\ref{eq:11}) is in fact not very good, and a much better
agreement with numerical simulations is obtained by performing
numerically the summation in the original discrete expression
Eq.~(\ref{eq:12}). This fact is due to the effects of the continuum
approximation in the index $i$, which are negligible at the level of
the degree distribution, but show up at the level of correlations.

Finally, for the average degree of the nearest neighbors of the
vertices of degree $k$, $\bar{k}_{nn}(k)$, we resort to the expression
Eq.~(\ref{eq:13}), taking the form for the SM
\begin{equation}
  \bar{k}_{nn}(k) = 1 + \frac{1}{N P(k) k!} \sum_{i=1}^N
  \exp[-\bar{k}(i)]\bar{k}(i)^k \bar{k}_{nn}(i). \label{eq:14}
\end{equation}
This expression is far too complex to obtain even an asymptotic
expression in the continuous $i$ approximation, so we will compare
numerical simulations with a direct numerical evaluation of the
summation in Eq.~(\ref{eq:14}).

\section{Numerical simulations}
\label{sec:numer-simul}

We have checked the analytical predictions presented in the previous
Section by means of extensive numerical simulations of the original
SM.  We have generated networks with $\alpha$ variable, $m=3$ and size
$N=10^{5}$. All results are averaged over $10^{3}$ realizations for
each value of the parameter $\alpha$.  Simulation were performed as
follows: At each iteration, we extract a pair of real numbers
according to a power-law probability distribution with exponent $\alpha$,
normalized between $0.5$ and $N+0.5$.  Number are extracted using the
Monte Carlo inversion method \cite{montecarlo}.  Then, we approximate
each number to the nearest integer, so that the resulting pair is
composed by integers between $1$ and $N$. These are the two candidate
vertices to be connected by and edge. If the proposed pair is composed
by two identical numbers, or it has been extracted before, the
extraction is rejected and repeated until two valid vertices are
proposed.  We iterate this procedure until a network of $E=mN$ edges
is created. This algorithm corresponds exactly to the original SM. The
only modification is that the probability distribution according to
which we extract the candidate edges is not discrete, but continuous.
Anyway, it is possible to see that the results of the proposed
procedure are indistinguishable from those obtained from methods that
start directly from a discretized distribution, but require more
computation time (for example, by using the rejection method
\cite{montecarlo}).

\begin{figure}
  \centerline{\epsfig{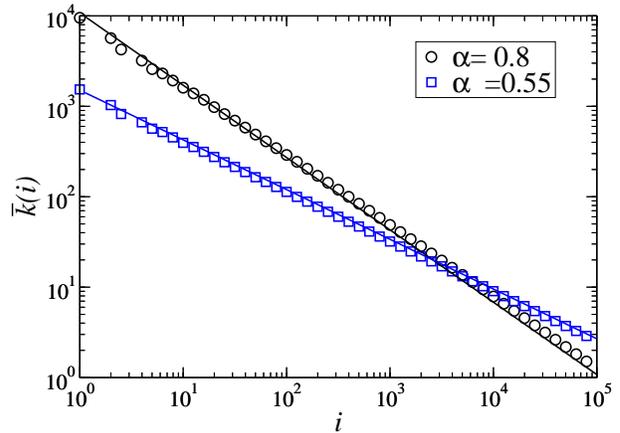}}
  \caption{Average degree of the vertices with index $i$ in the
    original SM, for two different values of $\alpha$. The solid lines
    represent the theoretical value given by
    Eq.~(\protect\ref{eq:9}).}
  \label{fig:csmKI}
\end{figure} 

\begin{figure}
  \centerline{\epsfig{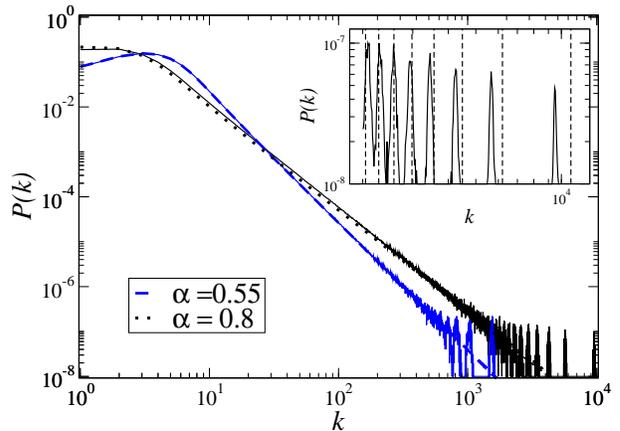}}
  \caption{Degree
    distribution in the SM for two different values of $\alpha$.  The
    dotted and dashed lines represent the theoretical value given by
    Eq.~(\protect\ref{eq:10}). In the inset: enlargement of the tail
    of the distribution. Peaks due to the discretization of the degree
    are visible. Dashed vertical lines represent the theoretical
    values of the centers of such peaks. }
  \label{fig:csmPK}
\end{figure} 

In Fig.~\ref{fig:csmKI} we plot the average degree of the vertices
with index $i$, $\bar{k}(i)$ for two different values of $\alpha$, namely
$\alpha=0.55$ and $\alpha=0.8$, which correspond to the degree exponents
$\gamma=2.82$ and $\gamma=2.25$, respectively. In both cases, the analytical
result, as as given by Eq.~(\ref{eq:9}), fit almost perfectly the
curves emerging from numerical simulation. The same happens for the
degree distribution, shown in Fig.~\ref{fig:csmPK}, for the two values
of $\alpha$ considered.
%In fact
%tail of the distribution could be fitted with a power-law, with an
%exponent in good agreement with the estimate of Eq.~(\ref{eq:degexp}).
As we can see from this Figure, the complete expression calculated in
Eq.~(\ref{eq:10}) fits exactly the whole distribution, except at very
large values of $k$. This discrepancy, due to the finite size of the
networks, is easy to understand. From Eq.(\ref{eq:9}), we can observe
that large values of $k$ correspond to small values of the index
$i$. In this region, the continuous $i$ and $k$ approximation made in
all calculations is expected to fail, and the index $i$ to show its
true discrete nature.  Indeed, this fact can be clearly observed in
the inset in Fig.~\ref{fig:csmPK}, where we plot a close-up of the
tail of the degree distribution obtained for $\alpha=0.8$, obtained from
averaging over $10^{3}$ network samples. This plot shows a set of peaks,
corresponding to the first values of the index $i$, from $1$ to $8$.
The centers of the peaks are well approximated by the analytical
$\bar{k}(i)$ function given in Eq.~(\ref{eq:9}), and represented by
means of vertical dotted lines, except for very small values of $i$.
The width of the peaks is accounted for by the fluctuations in the
value of $k$ in the different network samples.

\begin{figure}
  \centerline{\epsfig{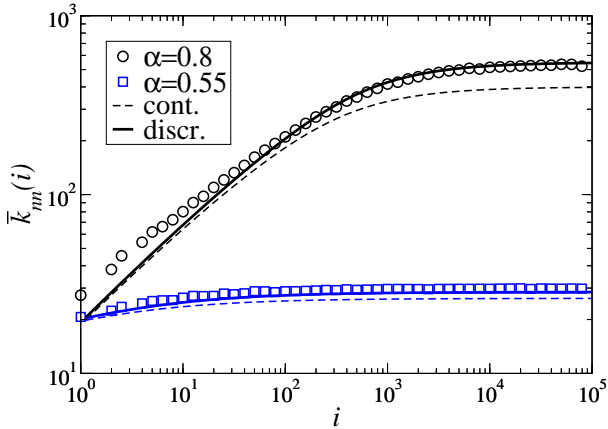}}
  \caption{Average nearest neighbor degree of the vertices with index
    $i$ in the SM for two different values of $\alpha$.  The
    solid lines represent the theoretical value given by the numerical
    summation of Eq.~(\protect\ref{eq:12}). The dashed lines correspond
    to the analytical approximation in the Eq.~(\protect\ref{eq:11}).}
  \label{fig:csm2}
\end{figure}

\begin{figure}
  \centerline{\epsfig{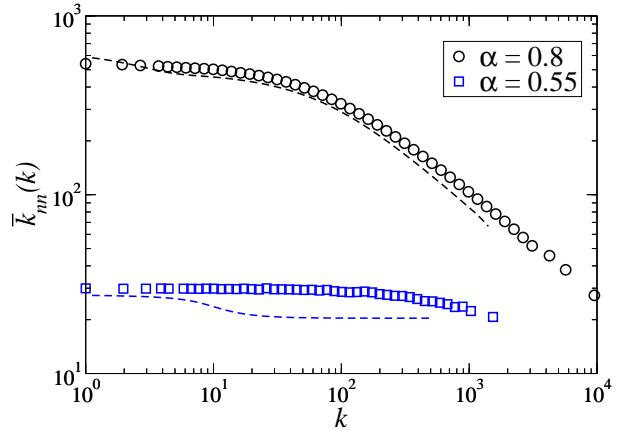}}
  \caption{Average nearest
    neighbor degree of the vertices with degree $k$, in the SM
    for two different values of $\alpha$.  The dashed lines correspond to the
    numerical summation of Eq.~(\protect\ref{eq:14}).}
  \label{fig:csm3}
\end{figure}

In Fig.~\ref{fig:csm2} we report the average nearest neighbors degree
of the vertices with index $i$. The dashed line represents the
theoretical approximation obtained in Eq.~(\ref{eq:11}).  We find a
percentually small difference between calculation and simulation. This
difference can be attributed to the effect of the continuous
approximation.  Indeed, if we numerically calculate the sum of
Eq.~(\ref{eq:12}) and report it in the plot (continuous line), we
obtain a better fit of the simulation results. A good agreement
between theory and simulation is obtained as well in the plot of the
average nearest neighbor degree of the vertices with degree $k$,
Fig.~\ref{fig:csm3}, at least for sufficiently large values of $\alpha$.
Here theoretical value is obtained directly from numerical summation
of Eq.~(\ref{eq:14}), that cannot be approximated analytically in a
simple way.  The correlation function displays an almost constant
behavior at low degrees and a decreasing slope at high degrees, i.e.
a regime without any correlation followed by one characterized by
strong disassortative mixing.  The emergence of these correlation is
connected to the absence of multiple and self-connections. By doing
this, we bias the natural tendency of high degree vertices to have
some connections into each other, favoring their linking to small
degree vertices, and, therefore, generating negative correlations in
the degree. This phenomenon appears to be extremely relevant for small
values of $\gamma$. In the case $\alpha=0.8$, for example, we can observe in the
average neighbor connectivity a decay of more than one decade in
about two decades of the degree. On the other hand, the analytical
solutions does not behave so well for large values of $\gamma$ (small $\alpha$),
probably due to the accumulated effect of all the approximations made
in obtaining this expression. 

\section{Conclusions}
\label{sec:conclusions}

In this work we have presented an analytic solution of the static
model \cite{static}, which has been recently proposed as a
paradigmatic scale-free non-growing network model.  The solution is
obtained via a mapping of the SM into a hidden variables network
model, that represents its canonical counterpart, i.e. in which the
number of edges is not held fixed, but whose average degree tends to a
constant in the infinite network size limit.  We have derived
analytically the properties of the mapped hidden variables network
model and checked the predictions by means of extensive numerical
simulations of the original model. The good agreement observed implies
that the canonical version of the SM is identical to the original
version in the thermodynamic limit $N\to\infty$.  It is particularly
noteworthy that our analytical calculations have allowed us to
evaluate the correlations induced in the model by the physical
condition of absence of self and multiple connections. The detected
amount of correlation is considerable and can have a strong influence
both on the topology of the networks, and on dynamics running on top
of them. The presence of these correlations thus casts some shadows on
the usefulness of the SM as a benchmark for mean-field solutions of
dynamical processes, which are usually obtained in the uncorrelated
limit.  In this sense, the recently proposed \emph{Uncorrelated
  Configuration Model} \cite{ucm} appears to be a more adequate
instrument for the investigation of the effects of scale-invariance in
network topology and dynamics, without the perturbations induced by
the presence of correlations.

\begin{acknowledgement}
  We thank M. Bogu\~n\'a for helpful comments and discussions. This work
  has been partially supported by EC-FET Open Project COSIN No.
  IST-2001-33555.  R.P.-S. acknowledges financial support from the
  Ministerio de Ciencia y Tecnolog\'\i a (Spain), and from the Departament
  d'Universitats, Recerca i Societat de la Informaci\'o, Generalitat de
  Catalunya (Spain). M. C. acknowledges financial support from
  Universitat Polit\`ecnica de Catalunya.
\end{acknowledgement}

\end{document}